\documentclass[11pt,a4paper]{article}
\usepackage{ifpdf}
\pdfoutput=1
\usepackage{amsmath}
\usepackage{amssymb}
\usepackage{mathrsfs}
\usepackage{color}
\usepackage{graphicx}
\usepackage[percent]{overpic}
\usepackage{url}
\usepackage{subfig}
\usepackage{enumerate}
\usepackage{overpic}
\usepackage{amsthm}
\usepackage{moreverb}
\usepackage{authblk}

\def\d{{\bm d}}

\def\n{{\bm n}}
\def\u{{\bm u}}
\def\v{{\bm v}}
\def\w{{\bm w}}
\def\x{{\bm x}}

\def\0{\boldsymbol{0}}
\def\ss{\boldsymbol{\sigma}}

\def\mubar{\overline{\mu}}
\def\ubar{\overline{\u}}
\def\vbar{\overline{\v}}

\def\qbar{\overline{q}}
\def\dt{\partial_t}

\def\cl {\nonumber \\}
\def\el {\nonumber }

\newtheorem{rem}{Remark}[section]

\newcommand{\bm}[1]{\mbox{\boldmath{$#1$}}}
\def\div{\nabla\cdot}

\usepackage{verbatim}
\usepackage{graphicx}
\usepackage{epstopdf}

\begin{document}
\date{}
\title{Fluid-structure interaction simulations with a LES filtering approach in \emph{solids4Foam}}


\author[1]{Michele Girfoglio\thanks{mgirfogl@sissa.it}}
\author[2]{Annalisa Quaini\thanks{quaini@math.uh.edu, ORCID ID 0000-0001-9686-9058}}
\author[1]{Gianluigi Rozza\thanks{grozza@sissa.it, ORCID ID 0000-0002-0810-8812}}
\affil[1]{SISSA, International School for Advanced Studies, Mathematics Area, mathLab, via Bonomea, Trieste 265 34136, Italy\\
}
\affil[2]{Department of Mathematics, University of Houston, Houston TX 77204, USA}
\maketitle

\begin{abstract}
The goal of this paper is to test solids4Foam, the
fluid-structure interaction (FSI) toolbox developed for foam-extend (a branch of OpenFOAM),
and assess its flexibility in handling more complex flows. For this purpose, we consider the interaction of an incompressible fluid described by a Leray model with a hyperelastic
structure modeled as a Saint Venant-Kirchhoff material. 
We focus on a strongly coupled, partitioned fluid-structure interaction (FSI) solver in a finite volume environment, combined
with an arbitrary Lagrangian-Eulerian approach to deal with the motion of the fluid domain. 
For the implementation of the Leray model, which features a nonlinear differential low-pass filter, we adopt a three-step algorithm called 
Evolve-Filter-Relax. We validate our approach against numerical data available in the literature for the 3D cross 
flow past a cantilever beam at Reynolds number 100 and 400.  


\end{abstract}
\vspace*{0.2cm}
\vspace*{0.5cm}

\section{Introduction}

Fluid–structure interaction (FSI) \cite{Hughes2013, Hou2012, Bungartz2006, Bungartz2010} 
involving incompressible fluid flows and flexible structures are found in a wide range of 
applications in both industrial and biomedical engineering. 
Thus, the search for accurate, robust, and efficient solvers has motivated a large body of literature. 
A fluid-structure problem is defined by a set of governing
equations to be fulfilled in the fluid domain and  a set of equations posed in the structure domain, plus suitable coupling conditions ensuring the
continuity of velocity (kinematic condition) and normal stress (dynamic condition) across the fluid-structure interface.  
One way to categorize algorithms for FSI problems is to divide them into \emph{partitioned} methods
(e.g., \cite{piperno.farhat.eal:partitionedI,fernandezg2,QQ2007,deparisd1})
and \emph{monolithic} methods (e.g., \cite{BATHE19991,Heil,BADIA20084216}). In partitioned methods, the solid and fluid domains are solved separately, by using 
one's favorite numerical methods and different, possibly non-matching computational grids. 
Partitioned methods can be further divided into \emph{strongly coupled} schemes
(e.g., \cite{causing1,gerbeauv1,fernandezm1}), 
which enforce the discrete counterpart of both coupling conditions (kinematic and
dynamic) up to a tolerance of choice, and \emph{weakly} or \emph{loosely coupled}
(e.g., \cite{piperno:explicit,burmanf2,BUKAC2013515}), 
for which the coupling conditions are not ``exactly'' satisfied at each time-step.
Notice that strongly coupled methods are generally stable in the energy norm. 
In monolithic approaches one deals with the coupled problem as one whole by adopting a unique,
ad-hoc numerical strategy. Monolithic methods are strongly coupled by design. 
Both families of methods have benefits and drawbacks, and the ``best'' choice 
mainly depends on the FSI problem under consideration. 
For more details, the reader is referred to, e.g., \cite{Hou2012, Degroote2009, Degroote2011, Richter2015, Slone2004}.

The goal of this paper is two-fold: i) to test solids4Foam \cite{Cardiff2018}, 
the advanced solid mechanics and FSI toolbox developed for foam-extend, a branch of 
OpenFOAM \textsuperscript{\textregistered} \cite{Weller1998}; and (ii) 
assess its flexibility in handling more complex flows. OpenFOAM \textsuperscript{\textregistered} is
an open source finite volume C++ library widely used by commercial and academic organizations.
Thus, the value of this paper lies in testing the current 
(at the time of writing this paper) release of a FSI library that is available to the scientific
community for free and thus is potentially used by many. In addition, we investigate the
capability of solids4Foam to handle more complex fluid models, since 
FSI problems with flows at higher Reynolds numbers have received less attention (some relevant references
are \cite{Breuer2012, Revstedt2013, Lorentzon2019}).
Towards our goal, we consider a strongly coupled, partitioned solver for the interaction between an incompressible fluid at 
moderately large Reynolds numbers and an elastic, compressible structure exhibiting ``large'' displacements and rotations. 
Our solver is based on a finite volume (FV) discretization method for both fluid and solid sub-problems. 

 
In the FSI problem we focus on, the fluid problem is given by a Leray model with a 
nonlinear differential low-pass filter in the arbitrary Lagrangian-Eulerian (ALE) 
formulation. 
Notice that by ``large'' structural displacement, we mean that non-negligible but not large enough to
to make the ALE solver crash. 
For the implementation of the Leray model, we adopt a three-step algorithm Evolve-Filter-Relax (EFR) \cite{BQV, Girfoglio2019}. 
To the best of our knowledge, it is the first time that the Leray model is used within a FSI context, although 
obviously other LES approaches have been used.  In particular, 
see \cite{Rege2017, Sekutkovski2015} for numerical results obtained with LES techniques implemented 
in foam-extend and applied to FSI problems.
The big advantage of the EFR method is modularity: its implementation does not
require any major modification of a legacy solver. 
The structure problem is given by compressible elasticity. In particular, the
solid is modeled as a St.~Venant-Kirchhoff hyperelastic material. 
To approximate the solution of the coupled FSI problem, we adopt the Dirichlet-Neumann (DN) method, i.e.~the 
fluid problem is endowed with a Dirichlet boundary condition at the fluid-structure interface, enforcing the
kinematic coupling condition, and the structure problem is supplemented 
with a Neumann boundary condition at the interface, enforcing the dynamic coupling condition. 
A weakly coupled DN method is unconditionally unstable when the added mass effect is large \cite{causing1}, i.e.~the 
densities of the fluid and solid are of the same order of magnitude. 
A strongly coupled DN method may require relaxation and, if the relaxation parameters are not suitably chosen, 
the convergence of the DN method could be slow. 
The particular  DN method we use in this work is called IQN-ILS, which stands for
interface quasi-Newton with inverse Jacobian from a least-squares model \cite{Degroote2010, Degroote2009}.

An important outcome of this work is that the code created for it is incorporated in an open-source library\footnote{\url{https://mathlab.sissa.it/cse-software}} and therefore is readily shared with the community.

In order to validate our solver, we consider a 3D flexible plate embedded in a cross flow at Reynolds 
number 100 and 400. We compare our results with those obtained in 
\cite{Tian2014}, where the authors use a finite difference based immersed-boundary method for the fluid 
problem and a finite element formulation for the structure problem. 
This benchmark has also been studied in laminar regime in \cite{Richter2011}, 
as well as at higher Reynolds numbers (of the order of a thousand) in \cite{Lorentzon2019, Rege2017, Zhu2011}.  

The outline of this paper is as follows. In Sec.~\ref{sec:math_fram}, we introduce the mathematical framework, 
including the governing equations of the fluid and the solid, and the coupling condition. 
In Sec.~\ref{sec:math_disc}, we detail our numerical strategy for time and space discretization. The numerical results for the elastic beam in a cross flow are reported in Sec.~\ref{sec:beam}. Finally, conclusions are drawn in Sec.~\ref{sec:conclusions}.

\section{Problem definition}\label{sec:math_fram}

We deal with the interaction between an incompressible Newtonian fluid at moderately large Reynolds number
and a hyperelastic solid.
The fluid model and the structure model are described in Sec.~\ref{sec:eq_fluids} and \ref{sec:eq_solids}, respectively. 
The coupling conditions are specified in Sec.~\ref{sec:eq_inter}. The ALE problem is stated in Sec.~\ref{sec:mesh_def}.

\subsection{Fluid subproblem}\label{sec:eq_fluids}

We consider the motion of an incompressible viscous fluid in a spatial domain $\Omega_f (t)$ 
whose shape is changing in time $t \in (t_0, T)$ due to the
deformation of the structure that covers part of the fluid domain boundary. 

In order to describe the evolution of the fluid domain, we adopt an \emph{Arbitrary 
Lagrangian-Eulerian} (ALE) approach, see, e.g. \cite{hughesl1}. Let $\Omega_0 \subset \mathbb{R}^d$ 
be a fixed reference domain, e.g. $\Omega_0 = \Omega_f(0)$ . We consider a smooth mapping
\begin{align}
&A_t: \Omega_0 \longrightarrow \Omega_f(t), \nonumber \\
&A_t:  \x_0 \mapsto \x, \nonumber
\end{align} 
where $\x$ and $\x_0$ are the coordinates 
in the physical domain $\Omega_f(t)$ and the reference domain $\Omega_0$, respectively.
For each time instant $t \in [t_0, T]$, $A_t$ is assumed to be a diffeomorphism. The domain velocity
$\boldsymbol{w}$ is given by 
\begin{align}
  \w(t, \cdot) = \frac{d A_t}{dt}(t, A_t(t, \cdot)^{-1}). \el
\end{align} 
The relationship between the rate of change of the volume $\Omega_f(t)$ and the velocity $\w$ is defined by the geometric (space) conservation law (GCL, see \cite{Thomas1997, Demirdizic1988}):
\begin{equation}\label{eq:CGL}
\dfrac{d\Omega_f}{dt} - \nabla \cdot \w = 0
\end{equation}
The ALE time derivative of a function $f(t, \x)$ is defined as:
\begin{equation}
\partial_t {f}|_{\x_0} = 
D_t f(t, A_t({\x}_0)) = \partial_t f(t, \x) + \w(t, \x) \cdot \nabla f(t, \x) \el, ~ \textrm{for} \; \x = A_t (\x_0),\; {\x}_0 \in \Omega_0.
\label{ALErelation}
\end{equation} 
Let $J_{A_t}$ denote the Jacobian of the 
deformation gradient, i.e. $J_{A_t}=\det\left(\frac{\partial \x}{\partial \x_0}\right)$. We have:
\begin{equation*}
\partial_t J_{A_t} |_{\x_0}=J_{A_t}\nabla \cdot \boldsymbol{w}.
\end{equation*} 

To describe the fluid motion, we adopt the so called \emph{Leray model}, which 
couples the Navier-Stokes equations 
with a differential filter.
With the above definitions, we can write the  Leray model
with the Navier-Stokes equations in the ALE formulation as follows:
\begin{align}
\rho_f\, \dt \left(J_{A_t} \u_f\right) |_{\x_0} + \rho_f\,J_{A_t}\div \left(\left(\ubar_f - \w\right) \otimes \u_f\right) - J_{A_t}\div \ss_f & = 0  \quad ~~{\rm in}~\Omega_f(t), \label{eq:filter-ns1}\\
\div \u_f &= 0  \quad ~~{\rm in}~\Omega_f(t), \label{eq:filter-ns2} \\
-2 \alpha^2\div \left(a(\u_f) \nabla\ubar_f\right) +\ubar_f +\nabla \lambda_f &= \u_f  \quad {\rm in}~\Omega_f(t) ,\label{eq:filter-mom}\\
\div \ubar_f &= 0  \quad ~~{\rm in}~\Omega_f(t),\label{eq:filter-mass}
\end{align}
for $t \in (t_0,T)$. 
Here,  $\u_f$ denoted the fluid velocity, $\ubar_f$ is the \emph{filtered velocity}, $p_f$ is the pressure, $\rho_f$ is the fluid density,
and $\mu_f$ is the constant dynamic viscosity.
For Newtonian fluids, the Cauchy stress tensor $\ss_f$ is given
\begin{equation}\label{eq:newtonian}
\ss_f (\u_f, p_f) = -p_f \mathbf{I} +\mu_f (\nabla\u_f + \nabla\u_f^T).
\end{equation}
In \eqref{eq:filter-mom}, $\alpha$ can be interpreted as the \emph{filtering radius} 
(that is, the radius of the neighborhood where the filter extracts information from the unresolved scales)
and the variable $\lambda_f$ is a Lagrange multiplier to enforce the incompressibility constraint for $\ubar_f$.

Scalar function $a(\cdot)$ is such that:
\begin{align*}
a(\u_f)\simeq 0 & \mbox{ where the velocity $\u_f$ does not need regularization;}\\
a(\u_f)\simeq 1 & \mbox{ where the velocity $\u_f$ does need regularization.}
\end{align*}
This function, called \emph{indicator function}, is crucial for the success of the Leray model. 
Different choices of $a(\cdot)$ have been proposed and compared in \cite{Borggaard2009,layton_CMAME,O-hunt1988,Vreman2004,Bowers2012}. 
A convenient indicator function is $a(\u) = | \nabla \u |$ (suitably normalized \cite{Borggaard2009}) 
because of its strong monotonicity properties. With this choice for the indicator function, 
we obtain a {\it Smagorinsky-like model}, which is however known 
to be not sufficiently selective. Indeed, it selects laminar shear flow (where $| \nabla \u|$ is
constant but large) as a region of the domain with turbulent fluctuations. Thus, we 
choose to work with a more selective class of indicator functions, which are based on 
the deconvolution operator. Such functions are defined as:
\begin{equation}
a(\u_f) = a_{D}(\u_f) = \left|  \u_f - D (F(\u_f)) \right|^2, \label{eq:a_deconv}
\end{equation}
where $F$ is a linear filter (an invertible, self-adjoint, compact operator from a Hilbert space to itself)
and $D$ is a bounded regularized approximation of $F^{-1}$. A popular choice for $D$ is the Van Cittert deconvolution operator $D_N$, defined as
\begin{equation}
D_N = \sum_{n = 0}^N (I - F)^n. \el
\end{equation}
In this paper  we consider $N = 0$, corresponding to $D_0=I$. For this choice of $N$, the indicator function (\ref{eq:a_deconv}) becomes
\begin{align}
a_{D_0}(\u_f) = \left|  \u_f - F(\u_f) \right|. \label{eq:a_D0_a_D1}
\end{align}
We select $F$ to be the linear Helmholtz filter operator $F_H$ defined by 
\begin{equation}
	F=F_H = \left(I - \alpha^2  \Delta \right)^{-1}. \el
\end{equation}
We highlight that finding $F_H(\u_f) = \tilde{\u}_f$ is equivalent to finding
$\tilde{\u}_f$ such that:
\begin{equation}\label{eq:vtilde}
\tilde{\u}_f - \alpha^2  \Delta \tilde{\u}_f = \u_f. 
\end{equation}

In order to characterize the flow regime under consideration, we define the Reynolds number as
\begin{equation}\label{eq:re}
Re = \frac{U b}{\nu},
\end{equation}
where $\nu=\mu/\rho$ is the \emph{kinematic} viscosity of the fluid, and $U$ and $b$ 
are characteristic macroscopic velocity and length, respectively. 

\subsection{Structure subproblem}\label{sec:eq_solids}
The deformation of the solid is assumed to be elastic and compressible. 
Let $\Omega_s(t)$ be the current solid domain, $\rho_s$ the solid density, 
and $\d_s$ is the structure displacement. The structure linear momentum conservation law
can be described by: 
\begin{equation}\label{eq:solid1}
\rho_s \partial_{tt} \d_s = \nabla \cdot \boldsymbol{\sigma_s} \quad {\rm in}~\Omega_s(t) 
\end{equation}
for $t \in (t_0,T)$. The Cauchy stress tensor $\boldsymbol{\sigma_s}$ is given by
\begin{equation}\label{eq:solid2}
\boldsymbol{\sigma_s} = \dfrac{1}{det \boldsymbol{F}} \boldsymbol{F} \cdot \boldsymbol{\Sigma} \cdot \boldsymbol{F}^T, 
\end{equation}
where $\boldsymbol{F} = \boldsymbol{I} + \nabla \d_s^T$ is the deformation gradient tensor, and $\boldsymbol{\Sigma}$ is the second Piola-Kirchhoff stress tensor. Let us also introduce the Green-Lagrange strain tensor
\begin{equation}\label{eq:solid3}
\boldsymbol{E} = \dfrac{1}{2} \left[\nabla \d_s + \nabla \d_s^T + \nabla \d_s \cdot \nabla \d_s^T \right].
\end{equation}

We consider the Saint Venant-Kirchhoff constitutive material model, for which
the second Piola-Kirchoff stress tensor and the Green-Lagrange strain tensor satisfy the following relationship:
\begin{equation}\label{eq:solid4}
\boldsymbol{\sigma_s} =  2\mu_s \boldsymbol{E} + \lambda_s tr(\boldsymbol{E}) \boldsymbol{I},
\end{equation}
where $\mu_s$ and $\lambda_s$ are the Lamè coefficients.
These coefficients are linked to the elastic modulus $E$ and the Poisson's ratio $\nu_s$ through:
\begin{align}\label{eq:E_nu}
\mu_s = \dfrac{E}{2\left(1 + \nu_s \right)},  \quad  \lambda_s = \dfrac{\nu_s E}{\left(1 + \nu_s \right)\left(1 - 2 \nu_s \right)}.
\end{align}

Notice that eq.~\eqref{eq:solid1} could be rewrite as follows
\begin{equation}\label{eq:solid5}
\rho_s \partial_{tt} \d_s - \nabla \cdot {(2\mu_s + \lambda_s) \nabla \d_s}= \nabla \cdot \boldsymbol{q}_s \quad {\rm in}~\Omega_s(t), \quad t \in (t_0,T),
\end{equation}
with
\begin{equation}\label{eq:solid6}
\boldsymbol{q}_s = \mu \nabla \d_s^T + \lambda tr (\nabla \d_s) \boldsymbol{I} - (\mu_s + \lambda_s) \nabla \d_s + \mu_s \nabla \d_s \cdot \nabla \d_s^T + \dfrac{1}{2}\lambda_s tr\left(\nabla \d_s \cdot \nabla \d_s^T\right)\boldsymbol{I} + \boldsymbol{\sigma_s} \cdot \nabla \d_s.
\end{equation}

\subsection{Coupling conditions}\label{sec:eq_inter}
The fluid and solid governing equations are coupled by the kinematic and dynamic coupling conditions at the 
fluid-structure interface $I(t)$. The kinematic condition ensures that the velocity is continuous across the interface: 
\begin{align}\label{eq:fsi_1}
\u_{f}  |_{I(t)} = \u_{s}  |_{I(t)}, \quad t \in [t_0, T].
\end{align}
The dynamic condition employs the equilibrium of the forces at the interface:
\begin{equation}\label{eq:fsi_2}
\n_I \cdot \boldsymbol{\sigma_f} = \n_I \cdot \boldsymbol{\sigma_s},
\end{equation}
where $\n_I$ is the unit normal vector at the interface. 


\subsection{ALE problem}\label{sec:mesh_def}
A classical choice for the displacement of the fluid domain $\d_f$ is to consider a 
harmonic extension of the structure displacement at the interface, i.e. 
\begin{align}
\nabla \cdot \left(\gamma \nabla \d_f \right) &= 0, \quad ~~\text{in}~\Omega_f(t), \label{eq:mesh_deformation}\\
\d_{f}  &= \d_{s}, \quad \text{on}~I(t). \label{eq:fsi_3}
\end{align}
for $t \in [t_0, T]$.
We opt for a variable diffusivity coefficient $\gamma$. In particular, we choose $\gamma$
to be inversely proportional to the square of distance from the moving boundary. 
Such a dependency proved to produce a smooth motion even 
in cases with moderate structure deformations \cite{Tukovic2012, Jasak2006}.

Notice that condition \eqref{eq:fsi_3} ensures that the fluid subdomain stays ``glued'' to the structure subdomain during the entire
time interval under consideration.

\section{Numerical approach}\label{sec:math_disc}

In this section, we report the details of the discretization of the fluid-structure interaction problem
described in Sec.~\ref{sec:math_fram}. 
First, in Sec.~\ref{sec:fsi_alg} we briefly describe the partitioned FSI algorithm we adopt. 
For the space discretization of all the subproblems ,
we choose the Finite Volume (FV) method \cite{Weller1998} that is derived directly from the integral form of the governing equations.
For the time discretization, we use Backward Differential Formula of order 1 (BDF1) \cite{quarteroni2007numerical}. 
Specifics of the discretization of the fluid and structure subproblems are presented in 
Sec.~\ref{sec:fluid} and \ref{sec:solid}, respectively. 
For discretization of ALE problem \eqref{eq:mesh_deformation}-\eqref{eq:fsi_3}, we refer the reader to \cite{Tukovic2018_bis}.

For the implementation, we chose the C++ finite volume library solids4Foam \cite{Cardiff2018}, the advanced solid mechanics and FSI toolbox developed for foam-extend, a branch of OpenFOAM \textsuperscript{\textregistered} \cite{Weller1998}.


\subsection{A partitioned FSI algorithm}\label{sec:fsi_alg}

The fluid-structure interaction problem is decoupled using the Dirichlet-Neumann (DN) procedure, 
where the flow problem is solved for a given velocity at the fluid-structure interface (Dirichlet interface condition), 
while the structural problem is solved for a given stress exerted on the interface (Neumann interface condition). 
Coupling conditions \eqref{eq:fsi_1}-\eqref{eq:fsi_2} are enforced at each time step 
through iterations between the fluid and solid solvers, i.e.~strong coupling.
Different options are available in solids4Foam to achieve strong coupling
with the DN scheme:
fixed relaxation \cite{causing1}, convergence acceleration using Aitken relaxation \cite{Kuttler2008}, 
and an interface quasi-Newton method with the approximation for the inverse of the Jacobian from a least-squares model (IQN–ILS) \cite{Degroote2010, Degroote2009}. The Aitken relaxation and the IQN–ILS procedures are preceded by two fixed-relaxation iterations. 
In this work, we use the IQN-ILS procedure.

The next two subsections are devoted to the discretization of the fluid and structure subproblems. 
We will denote the time step with $\Delta t \in \mathbb{R}$. 
Moreover, we will denote by $y^n$ the approximation of a generic quantity $y$ at the time $t^n= t_0 + n \Delta t$, with $n = 0, ..., N_T$ and $T = t_0 + N_T \Delta t$.

\subsection{Discretization of the fluid subproblem}\label{sec:fluid}

For the time discretization of problem (\ref{eq:filter-ns1})-(\ref{eq:filter-mass}), we adopt the Backward Euler (or BDF1) scheme.
To decouple the Navier-Stokes system (\ref{eq:filter-ns1})-(\ref{eq:filter-ns2}) from the filter system (\ref{eq:filter-mom})-(\ref{eq:filter-mass}) at each time step, we consider the Evolve-Filter-Relax (EFR), which was first proposed in \cite{layton_CMAME}. 
This algorithm reads as follows: given the Jacobians $J^{n}$ and $J^{n-1}$ of
ALE maps $A_{t^{n}}$ and $A_{t^{n-1}}$, the velocities $\u^{n-1}$ and $\u^{n}$ and domain $\Omega_f^n$ at $t^{n+1}$:

\begin{itemize}
\item[-] \textit{Evolve}: find intermediate velocity and pressure $(\v_f^{n+1},q_f^{n+1})$ such that
\begin{align}
\rho_f\, \frac{J^{n}\v_f^{n+1} - J^{n-1}\u_f^{n}}{\Delta t} &+ \rho_f\, \div \left(J^{n-1}\left(\u_f^n - \w^n\right)\otimes J^{n}\v_f^{n+1}\right) \cl
& - 2\mu_fJ^{n}\Delta\v_f^{n+1} +J^{n}\nabla q_f^{n+1}  = \boldsymbol{0} ~~{\rm in}~\Omega_f^n,\label{eq:evolve-1.1}\\
& \hspace{2.8cm} \div \v_f^{n+1}  = 0~~{\rm in}~\Omega_f^n \label{eq:evolve-1.2}.
\end{align}

\item[-] \textit{Filter}: find $(\vbar_f^{n+1},\lambda_f^{n+1})$ such that
\begin{align}
-\alpha^2\div \left(a(\v_f^{n+1}) \nabla\vbar_f^{n+1}\right) +\vbar_f^{n+1} +\nabla \lambda_f^{n+1} & = \v_f^{n+1}~~{\rm in}~\Omega_f^n, \label{eq:evolve-2.1}\\
\div \vbar_f^{n+1} & = 0\quad\quad{\rm in}~\Omega_f^n \label{eq:filter-1.2}.
\end{align}
\item[-] \textit{Relax}: set 
\begin{align}
\u_f^{n+1}&=(1-\chi)\v_f^{n+1} + \chi\vbar_f^{n+1}, \label{eq:relax-1} \\
p_f^{n+1}&= q_f^{n+1},  \label{eq:relax-2}
\end{align}
where $\chi\in(0,1]$ is a relaxation parameter.
\end{itemize}

\begin{rem}\label{rem:gen_Stokes}
Filter problem \eqref{eq:evolve-2.1}-\eqref{eq:filter-1.2} can be considered 
a generalized Stokes problem. In fact, by multiplying by $\rho_f$ and dividing 
by $\Delta t$ all the terms in eq.~\eqref{eq:evolve-2.1}, and rearranging the terms we obtain:
\begin{align}
\frac{\rho_f}{\Delta t} \vbar_f^{n+1}  - \div \left( \mubar_f \nabla\vbar_f^{n+1}\right) + \nabla \qbar_f^{n+1} & = \frac{\rho_f}{\Delta t} \v_f^{n+1}, \quad \mubar_f = \rho_f \frac{\alpha^2}{\Delta t} a(\v_f^{n+1}), \label{eq:filter-1.1}
\end{align}
where $\qbar_f^{n+1} = \rho_f \lambda^{n+1}/\Delta t$. Problem \eqref{eq:filter-1.1},\eqref{eq:filter-1.2} can be
seen as a time dependent Stokes problem with a non-constant viscosity $\mubar_f$, discretized
by the BDF1 scheme. 
\end{rem}

\begin{rem}
The EFR method has two appealing advantages over other LES models:
(i) it is modular, i.e.~it adds a differential problem to the Navier-Stokes problem
instead of extra terms in the Navier-Stokes equations themselves (like, e.g., the popular
variational multiscale approach \cite{BAZILEVS2007173}); 
(ii) the filter problem can be solved with a legacy Navier-Stokes solver, 
as shown in Remark \ref{rem:gen_Stokes}. Thus, thanks to the ERF method
anybody with a Navier-Stokes solver could simulate higher Reynolds number
flows without major modifications to the software core. For a thorough validation 
of the EFR method for Reynolds numbers up to 6500 in fixed domains, we refer to \cite{BQV, Girfoglio2019}.
\end{rem}

\begin{rem}\label{rem:p_relax}
In the EFR algorithm proposed in \cite{layton_CMAME} there is no relaxation for the pressure, i.e.~the end-of-step
pressure is set equal to the pressure of the Evolve step. 
In \cite{BQV}, two relaxations for the pressure were considered: $p^{n+1} = q_f^{n+1} + \gamma \chi \qbar^{n+1}$, where
$\gamma$ is a parameter related to the time discretization scheme (e.g., $\gamma = 1$ for BDF1), or
$p^{n+1} =(1-\chi)q_f^{n+1} + \chi\qbar_f^{n+1}$. Notice that while $\qbar_f^{n+1}$ has the same dimensional 
units as $q_f^{n+1}$, $\lambda_f^{n+1}$ does not.
\end{rem}

Next, we describe our choices for the space discretization. 
We partition the fluid computational domain $\Omega_f$ into moving cells or control volumes ${\Omega_f}_i (t)$,
with $i = 1, \dots, N_{f}$, where $N_{f}$ is the total number of cells in the fluid mesh. 
Let  \textbf{A}${_f}_j (t)$ be the surface vector of each face of the moving control volume, 
with $j = 1, \dots, M_f$. In order to simplify the notation and make equations more readable, 
in the following we will omit the subscript $f$ from most symbols. 


The fully discretized form of problem \eqref{eq:evolve-1.1}-\eqref{eq:evolve-1.2} is given by 
\begin{align}
&\rho_f\, \frac{\v^{n+1}_i}{\Delta t}\, \Omega_i^{n+1} + \rho_f\, \sum_j^{} \left[\left(\u^n_j - \w^n_j\right) \cdot \textbf{A}_j^{n+1} \right] \v^{n+1}_{i,j} \cl
& \quad \quad- 2\mu_f \sum_j^{} (\nabla\v^{n+1}_i)_j \cdot \textbf{A}_j^{n+1} + \sum_j^{} q^{n+1}_{i,j} \textbf{A}_j^{n+1}  = \rho_f\, \frac{\u^{n}_i}{\Delta t} \Omega_i^{n} \label{eq:disc_evolve1} \\
& \hspace{4cm} \sum_j^{} (\nabla q^{n+1})_j \cdot \textbf{A}_j^{n+1} = \sum_j^{} (\textbf{H}(\v_i^{n+1}))_j \cdot \textbf{A}_j^{n+1}, \label{eq:disc_evolve2}
\end{align}
where:
\begin{align}
\textbf{H}(\v^{n+1}_i) = -\rho_f \sum_j^{} \left(\u^n_j \cdot \textbf{A}_j^{n+1} \right) \v^{n+1}_{i,j} + 2\mu_f \sum_j^{} (\nabla\v^{n+1}_i)_j \cdot \textbf{A}_j^{n+1} + \rho_f\, \frac{\u^{n}_i}{\Delta t} \Omega_i^{n}. \label{eq:H} 
\end{align}
In \eqref{eq:disc_evolve1}-\eqref{eq:H}, $\v^{n+1}_i$ and $\u^{n}_i$ denotes the velocity and source term at the centroid of the moving control volume $\Omega_i^{n+1}$ and $\Omega_i^{n}$, respectively. Moreover, we denote with $\v^{n+1}_{i,j}$ and $q^{n+1}_{i,j}$ the velocity and pressure
associated to the centroid of face $j$. We remark that $\w^n_j$ satisfies the time-discretize version of eq. \eqref{eq:CGL}.

The fully discrete problem associated to the filter problem \eqref{eq:filter-1.1},\eqref{eq:filter-1.2} is given by 
\begin{align}
\frac{\rho_f}{\Delta t} \vbar^{n+1}_i - \sum_j^{}\mubar_j^{n+1}(\nabla\vbar^{n+1}_i)_j \cdot \textbf{A}_j^{n+1} +  \sum_j^{} \qbar^{n+1}_{i,j} \textbf{A}_j^{n+1} &= \frac{\rho_f}{\Delta t} \v^{n+1}_i, \label{eq:disc_filter1} \\
\sum_j^{} (\nabla \qbar^{n+1}_i)_j \cdot \textbf{A}_j^{n+1} &= \sum_j^{} (\overline{\textbf{H}}(\vbar^{n+1}_i))_j \cdot \textbf{A}_j^{n+1}, \label{eq:disc_filter2}
\end{align}
with
\begin{align}
\overline{\textbf{H}}(\vbar^{n+1}_i) =  \sum_j^{}\mubar_j^{n+1}(\nabla\vbar^{n+1}_i)_j \cdot \textbf{A}_j^{n+1} + \dfrac{\rho_f}{\Delta t}\v^{n+1}_i. \label{eq:Hbar}
\end{align}
In \eqref{eq:disc_filter1}-\eqref{eq:Hbar}, we denoted with $\vbar^{n+1}_i$
the filtered velocity at the centroid of the moving control volume $\Omega_i^{n+1}$, 
while $\qbar^{n+1}_{i,j}$ and $\mubar_j^{n+1}$ are the auxiliary pressure and artificial viscosity at the centroid of face $j$.  

Finally, the approximation of the problem \eqref{eq:vtilde} (which is needed to estimate the indicator function
\eqref{eq:a_deconv}) yields
\begin{align}\label{eq:disc_F}
\tilde{\v} ^{n+1}_i - \alpha^2 \sum_j^{} (\nabla \tilde{\v} ^{n+1}_i)_j \cdot \textbf{A}_j^{n+1} = \v^{n+1}_i,
\end{align} 
where $\tilde{\v} ^{n+1}_i$ is the value of $\tilde{\v} ^{n+1}$ at the centroid of the control volume $\Omega_i^{n+1}$. 

For more details related to the discretization of the Leray model, we refer the reader to \cite{Girfoglio2019,Girfoglio2020}. 
For a comprehensive description of space discretization with dynamic meshes, see \cite{Tukovic2012}.

For the solution of the linear system associated with \eqref{eq:disc_evolve1}-\eqref{eq:disc_evolve2} we used the PISO algorithm \cite{PISO}, while for problem \eqref{eq:disc_filter1}-\eqref{eq:disc_filter2} we chose a slightly modified version of the SIMPLE algorithm
\cite{SIMPLE}, called SIMPLEC algorithm \cite{Doormaal1984}. Both PISO and SIMPLEC are partitioned algorithms that decouple the computation of the pressure from the computation of the velocity. 

\subsection{Structure subproblem}\label{sec:solid}


We start by writing the time discretization of solid problem \eqref{eq:solid5}: given the displacements $\d_s^n$  and $\d_s^{n-1}$, 
find displacement $\d_s^{n+1}$ such that:
\begin{equation}\label{eq:disc_solid_time}
\rho_s \frac{\d_s^{n+1} - 2\d_s^n + \d_s^{n-1}}{\Delta t^2} - \nabla \cdot {(2\mu_s + \lambda_s) \nabla \d_s^{n+1}}= \nabla \cdot \boldsymbol{q}_s^{n} \quad {\rm in}~\Omega_s.
\end{equation}
where we recall that variabile $\boldsymbol{q}_s$ is defined in eq. \eqref{eq:solid6}.

Concerning the space discretization, we partition the initial undeformed computational solid domain $\Omega_s$ into cells or control volumes $\Omega_{s_i}$,
with $i = 1, \dots, N_{s}$, where $N_{s}$ is the total number of cells in the solid mesh. 
Let  \textbf{A}$_j$ 
be the surface vector of each face of the control volume, 
with $j = 1, \dots, M_s$. In an effort to keep notation simple, 
in the following we will omit the subscript $s$ from some variables. 

Then, the fully discretized form of problem \eqref{eq:solid5} is given by:
\begin{equation}\label{eq:disc_solid}
\rho_s \frac{\d_i^{n+1} - 2\d_i^n + \d_i^{n-1}}{\Delta t^2} {\Omega}_i - \left(2\mu_s + \lambda_s \right)\sum_j^{} (\nabla {\d} ^{n+1}_i)_j \cdot \textbf{A}_j = \sum_j^{} (\boldsymbol{q} ^{n}_i)_j \cdot \textbf{A}_j,
\end{equation}
where $\d_i^{n+1}$ denotes the displacement at the centroid of the control volume $\Omega_i$
and $(\boldsymbol{q} ^{n}_i)_j$ is the value of the variabile $\boldsymbol{q}$ 
associated to the centroid of face $j$.

We note that since we use a Lagrangian formulation for the solid problem (i.e. the momentum equation is integrated over the initial, undeformed configuration) the solid mesh is always in its initial configuration. 
For more details related to the discretization of the solid model, we refer the reader to, e.g., to \cite{Tukovic2018_bis}.

\section{Results}\label{sec:beam}

In this section, we present the numerical results aimed at validating our approach.
We consider a slender 3D flexible structure embedded in a cross flow at Reynolds number 100 and 400. 
Our simulation results will be compared against the numerical results from \cite{Tian2014}, which 
are referred as \emph{true} solutions hereinafter. 
We note that in \cite{Tian2014} a very different approach is used: 
a finite difference based immersed boundary method for the fluid flow and 
a finite element solver for the solid.
 This benchmark has been experimentally investigated at Reynolds number 1600
 for the first time in \cite{Luhar2011}, whose focus is the deformation of aquatic plants 
 in a water flow. 
 The authors of  both \cite{Tian2014} and \cite{Lorentzon2019} have used this
 benchmark to valide their FSI solver. 
The reason why we limit our investigation to Reynolds numbers 100 and 400
will be made clear in what follows.

\begin{figure}[h]
\centering
\subfloat[][]{\includegraphics[width=0.47\textwidth]{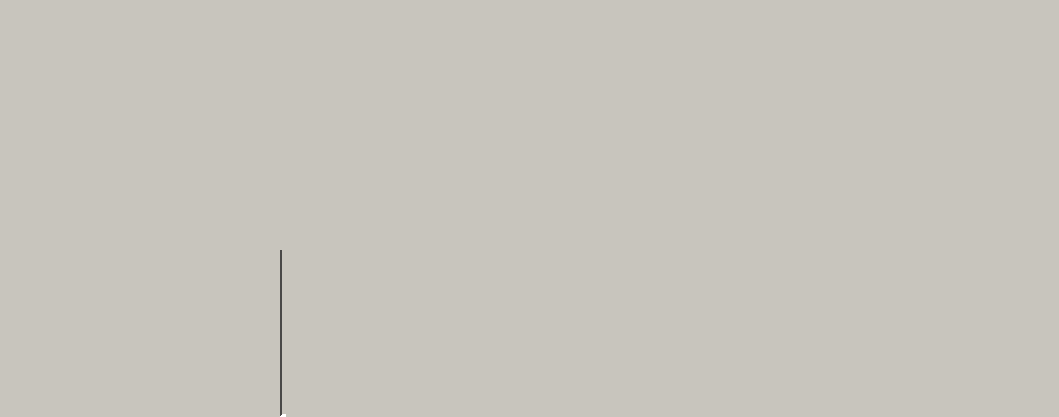}}~
\subfloat[][]{\includegraphics[width=0.26\textwidth]{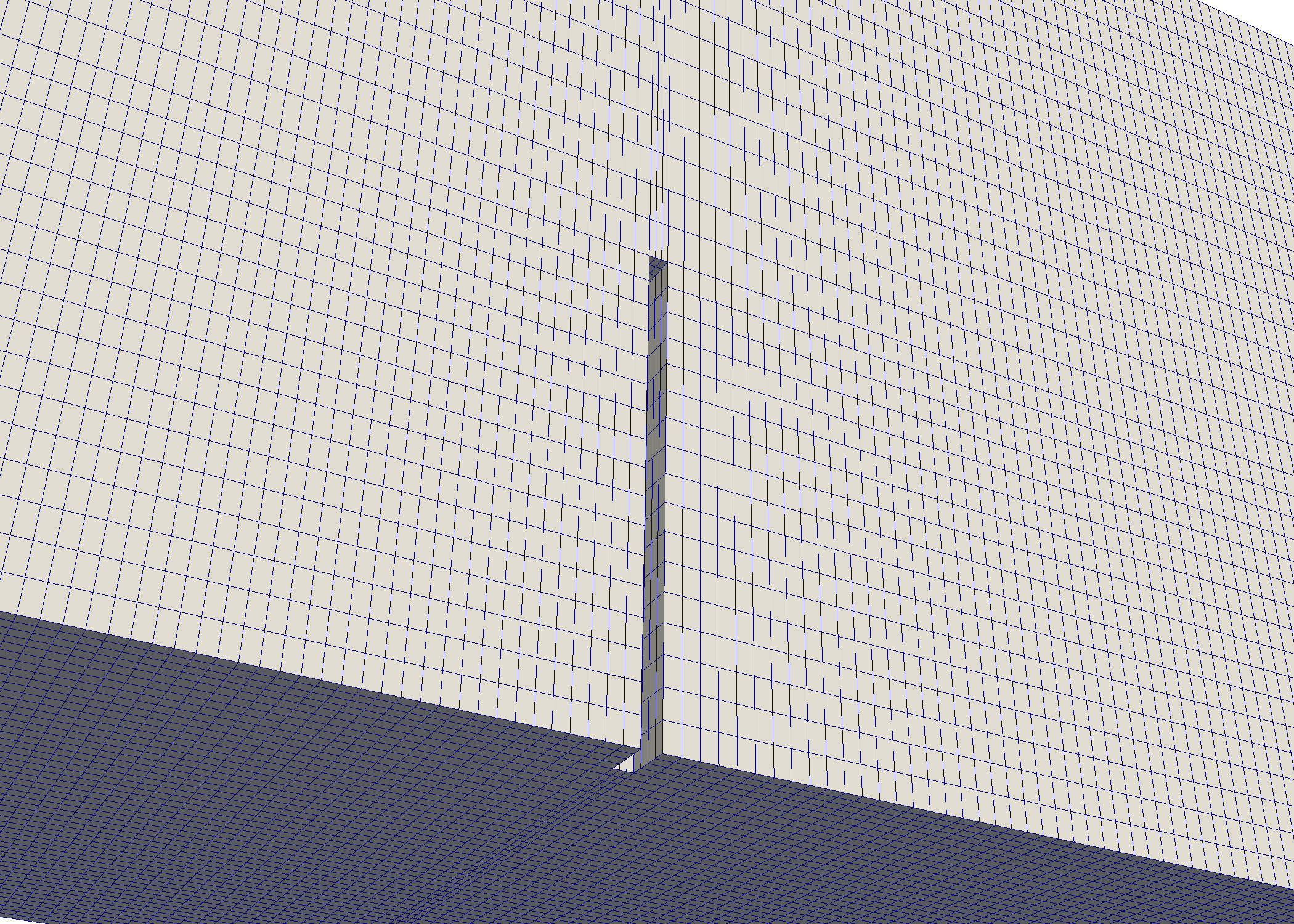}}~
\subfloat[][]{\includegraphics[width=0.22\textwidth]{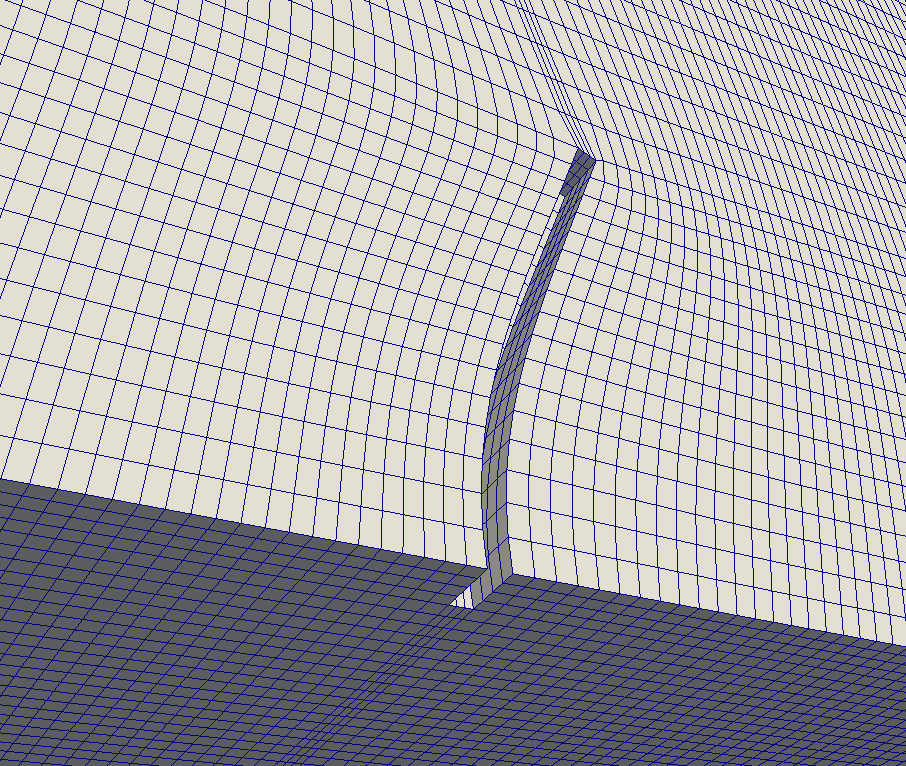}}
\caption{(a) Sketch of computational domain, (b) part of the undeformend fluid mesh $300k$ and c) one deformed
configuration of the same mesh.}
\label{fig:geometry}
\end{figure}

The computational domain is a 0.35 m $\times$ 0.125 m $\times$ 0.075 m
parallelepiped channel 
with an immersed structure of length $L = 0.05$ m, width $b = 0.01$ m, and thickness $h = 0.002$ m. 
See Fig.~\ref{fig:geometry} (a).
One of the structure ends is clamp-mounted, while the other is free. 
The structure is located initially at $x = 0.1$ m at the bottom of the channel. 
At the Reynolds numbers under consideration, we expect the flow pattern, and therefore the structure deformation, 
to be symmetric. Thus, we roughly halve the computational domain used in \cite{Tian2014}.  
This choice is also dictated by restrictions on the computations; see Remark \ref{rem_1}. 

Concerning the description of the flow, we consider both the Navier Stokes equations (NSE), 
i.e.~with no LES modeling, and the Leray model implemented through the EFR algorithm.
We impose a no slip boundary condition on the channel walls. 
As for the pressure at the channel walls, we calculate it 
using a linear extrapolation from the neighboring cell centre.
At the inflow, we prescribe a plug flow with horizontal velocity component $U$ at the centerline. 
At the outflow, we prescribe a null shear stress condition. 
We start all the simulations from fluid at rest.


We use the following parameters for our simulations: 
$\rho_s = 678$ Kg/m$^3$, $\nu_s = 0.4$, $E = 1905.49$ Pa, 
$U = 0.01$ m/s, $\rho_f = 1000$ Kg/m$^3$,  and $\mu_f = 1e-3$ Pa$\cdot$s 
to achieve $Re = 100$ or $\mu_f = 2.5e-4$ Pa$\cdot$ s to achieve $Re = 400$. 
These values are taken from \cite{Tian2014}, 
where they are stated in dimensionless form though.
The quantities of interest for this benchmark are 
the dimensionless horizontal ($\Delta x/b$) and vertical displacement ($\Delta y/b$)
of the midpoint of the structure free edge, 
and the drag coefficient:
\begin{align}\label{eq:cd_cl}
c_d(t) = \dfrac{2}{\rho_f b L {U}^2} \int_S \left(\boldsymbol{\sigma}_f
\cdot \boldsymbol{n}\right) \cdot \boldsymbol{t}~dS,
\end{align}
where $S$ is the structure surface, and $\boldsymbol{t}$ and $\boldsymbol{n}$ are the tangential and outward normal 
unit vectors, respectively. Since at $Re = 100, 400$ the flow evolves towards a steady state, 
these quantities are computed when the simulation is close enough to steady state  \cite{Tian2014}.
Moreover, for all simulations we evaluate the following errors:
\begin{align}\label{eq:Err}
E_{c_d} = \dfrac{c_{d} - c_{d}^{{true}}}{c_{d}^{{true}}}, \quad
E_{\Delta x/b} = \dfrac{\Delta x/b - {\Delta x/b}^{{true}}}{{\Delta x/b}^{{true}}}, \quad
E_{\Delta y/b} = \dfrac{\Delta y/b - {\Delta y/b}^{{true}}}{{\Delta y/b}^{{true}}}. \quad
\end{align}

We consider two different (initially) Cartesian orthogonal meshes for the fluid domain, 
while we consider only one mesh for the solid domain. 
The meshes are built by using \emph{blockMesh}, a mesh generation utility provided in OpenFOAM. 
Table \ref{tab:mesh} reports minimum and maximum diameter  in the initial configuration, 
and number of cells for each mesh. 
Fig.~\ref{fig:geometry} (b) and (c) show part of the fluid mesh $300k$ in the undeformed configuration
and one deformed configuration, respectively. 
We set the time step to 0.01 for all the simulations.

\begin{table}[htb!]
\centering
\begin{tabular}{|c|ccc|}
\hline
 & $h_{min}$ & $h_{max}$ & No. of cells  \\
\hline
Fluid mesh $300k$      & 6.67e-4  & 4.4e-3  & 364008       \\
\hline
Fluid mesh $700k$       &  6.67e-4  &  3.6e-3  &  695940   \\
\hline
Structure mesh      &   6.67e-4 & 2.4e-3   & 252   \\
\hline
\end{tabular}
\caption{Minimum diameter $h_{min}$ and maximum diameter $h_{max}$ in the initial configuration, and number of cells for all the meshes 
under consideration.}
\label{tab:mesh}
\end{table}

A Direct Numerical Simulation (DNS) needs a mesh with spacing close to the Kolmogorov scale $\eta$ \cite{Kolmogorov41-1,Kolmogorov41-2} that can be expressed as follows:
\begin{equation}\label{eq:eta-re}
\eta=Re^{-3/4}b.
\end{equation}
For our benchmark, we have $\eta = 3.16e-4$ at $Re = 100$ and $\eta = 1.12e-4$ at $Re = 400$. 
The meshes in Table \ref{tab:mesh} do not feature this level of refinement required by a DNS (especially for $Re = 400$), thus the need for LES modeling. 
We note that the fluid mesh used for the results in \cite{Tian2014} has about 
$6000k$ elements with $h_{min}^f = 2e-4$ and local refinement such that $h_{min}^f \sim \eta$ for 
all the Reynolds number under consideration. 
In fact, the results in \cite{Tian2014} are obtained with DNS.
We choose the under-refined meshes in Table \ref{tab:mesh} to be able to observe the effect of the filter.
As the mesh gets finer and finer (i.e., the mesh size gets closer to the Kolmogorov scale), 
all the relevant scales become resolved and no filter is needed. The reader interested
in learning about the performance of our LES approach for different levels
of mesh refinement in \emph{fixed} domains is referred to \cite{BQV, Girfoglio2019}.

\begin{figure}[htb!]
\centering
 \begin{overpic}[width=0.45\textwidth]{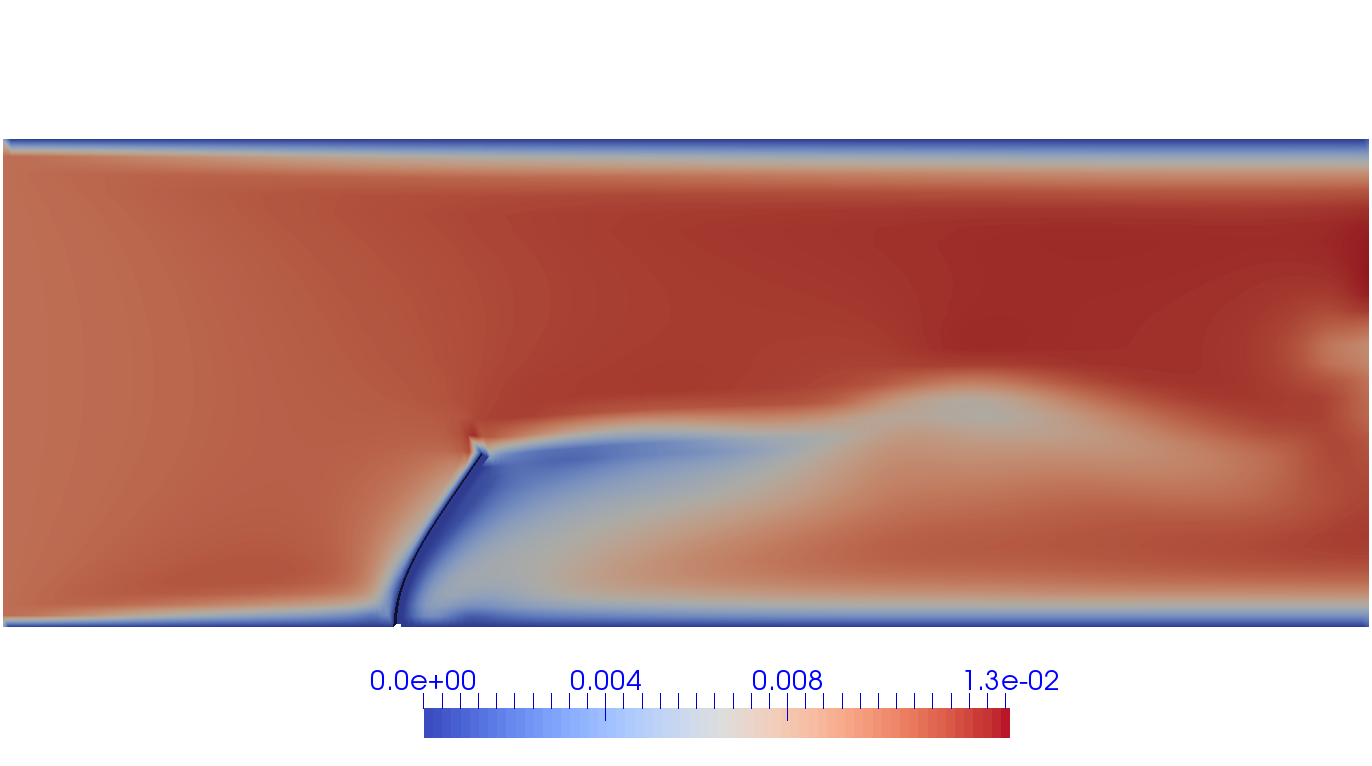}
      \end{overpic}
\begin{overpic}[width=0.45\textwidth]{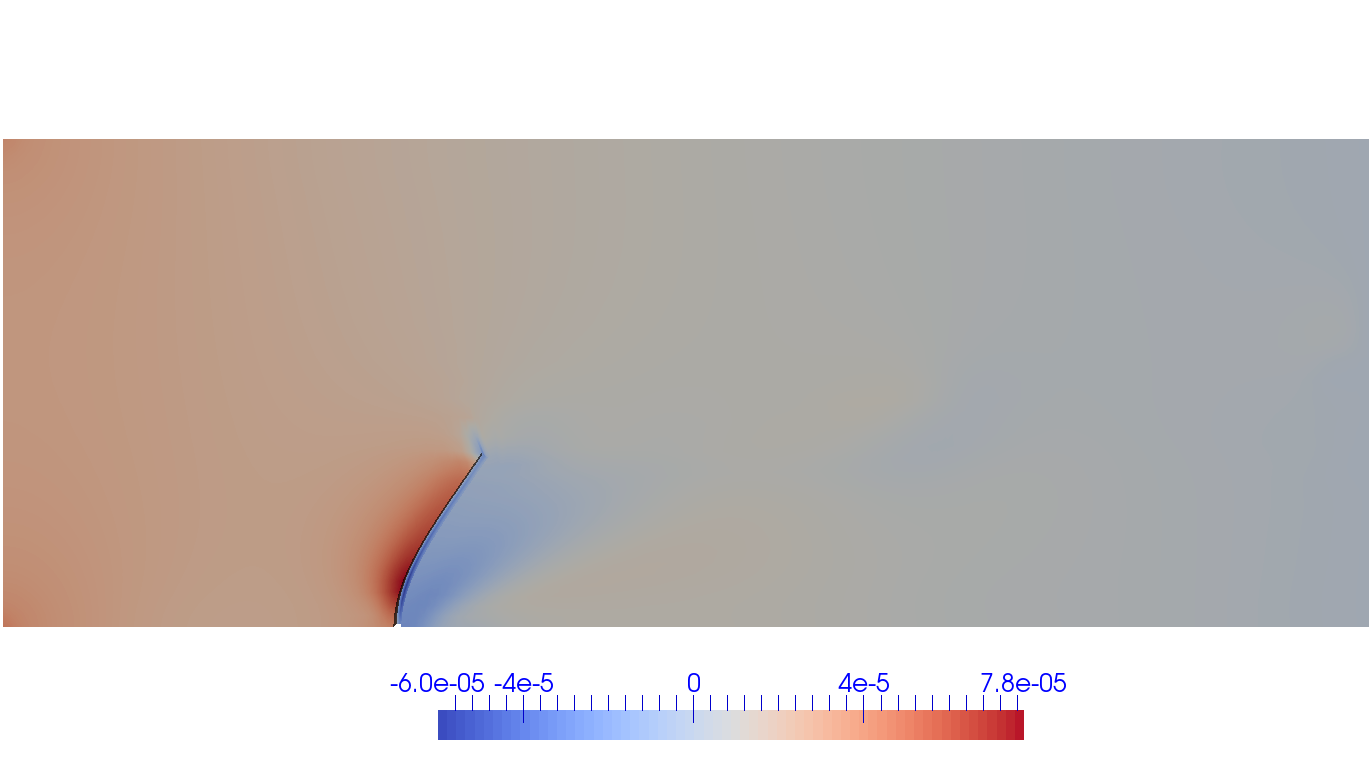}
      \end{overpic}\\
\begin{overpic}[width=0.45\textwidth]{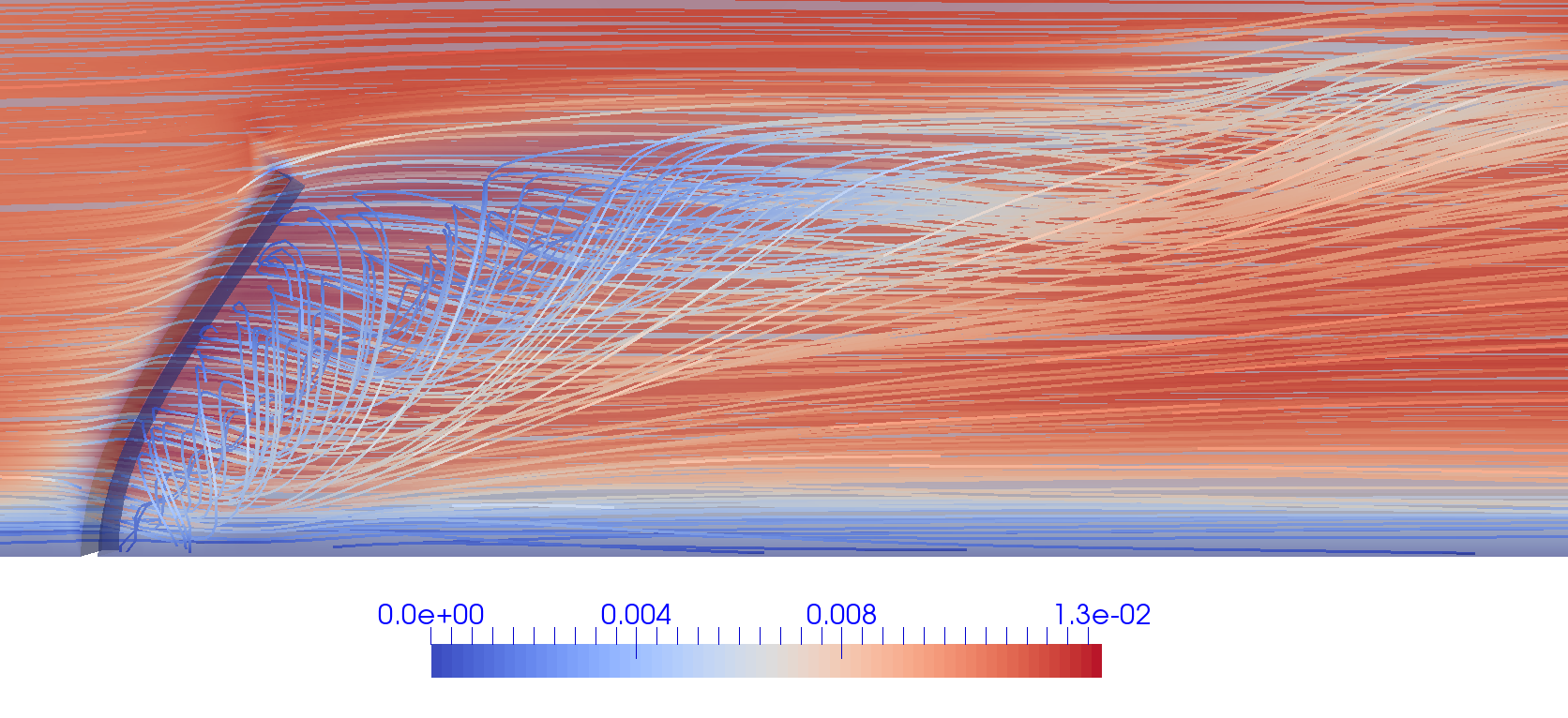}
      \end{overpic}
\begin{overpic}[width=0.45\textwidth]{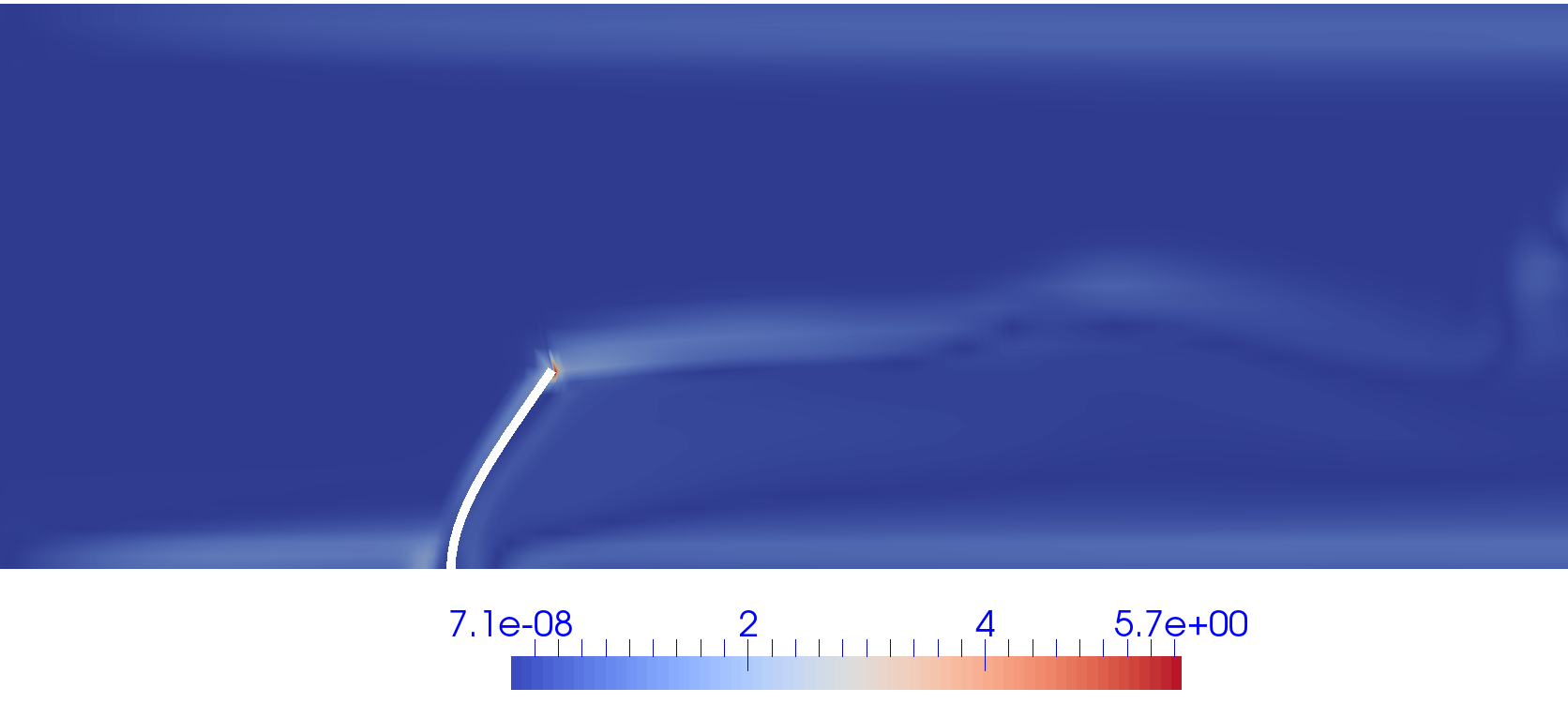}
      \end{overpic}\\
\caption{Mesh $300k$: Velocity $\u_f$ (top left), pressure $p_f$ (top right), streamlines (bottom left) and vorticity (bottom right) fields computed by the NSE algorithm 
close to steady state for Reynolds number 100.}
\label{fig:NSE}
\end{figure}

\begin{figure}[htb!]
\centering
 \begin{overpic}[width=0.45\textwidth]{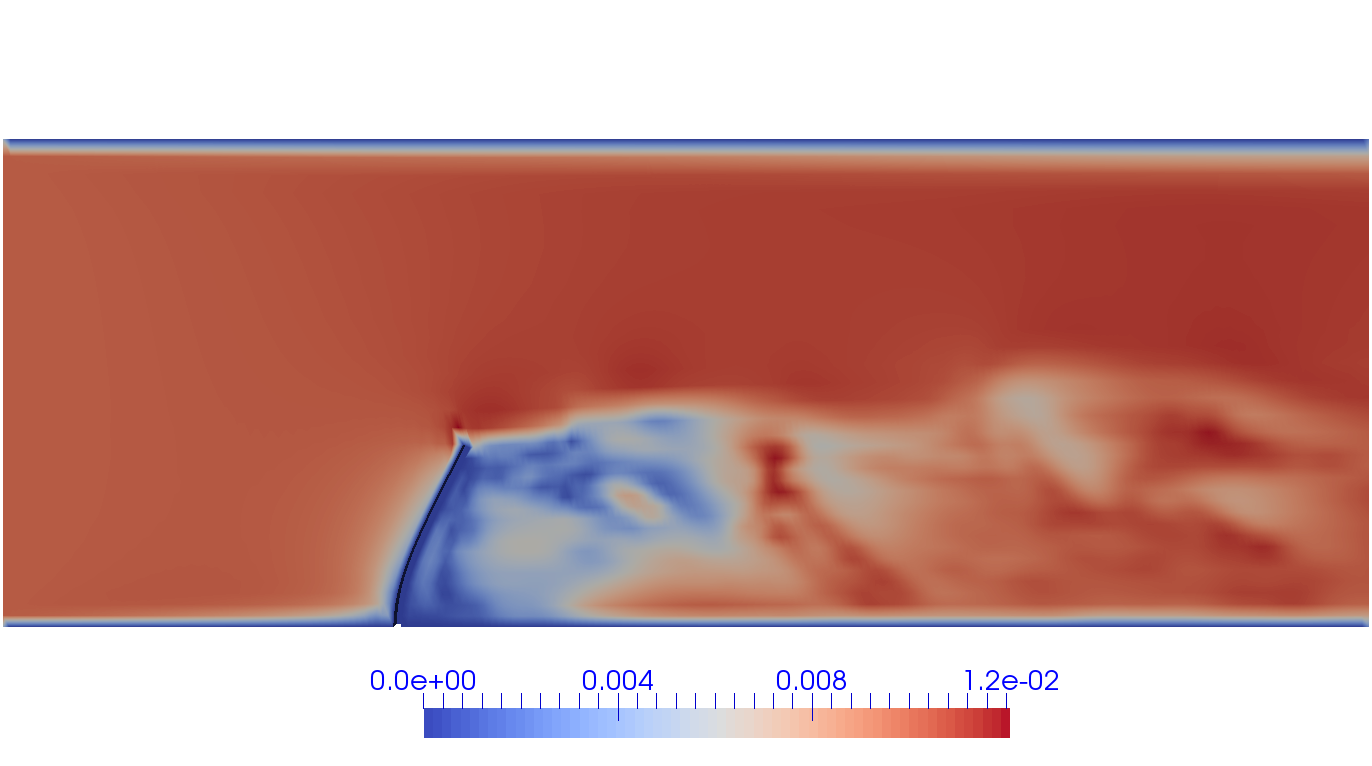}
      \end{overpic}
\begin{overpic}[width=0.45\textwidth]{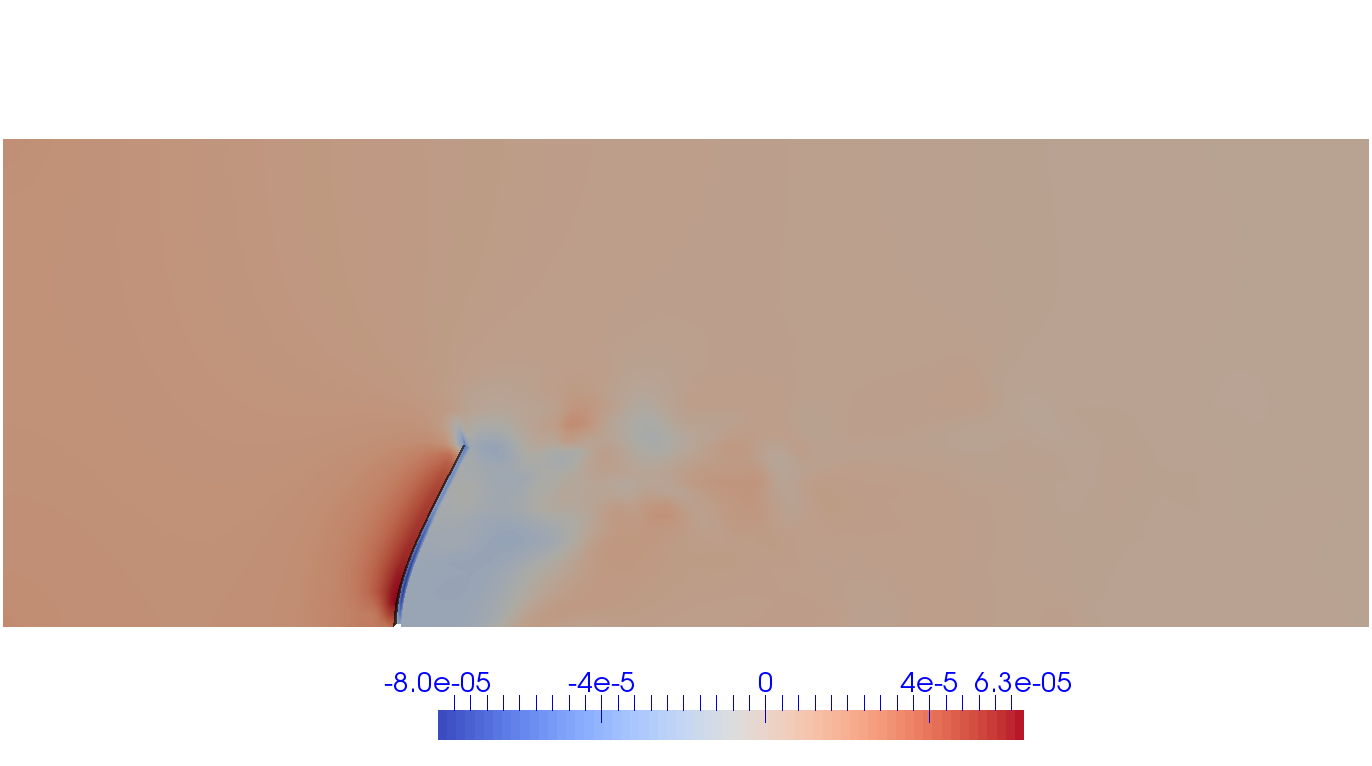}
      \end{overpic}\\
\begin{overpic}[width=0.44\textwidth]{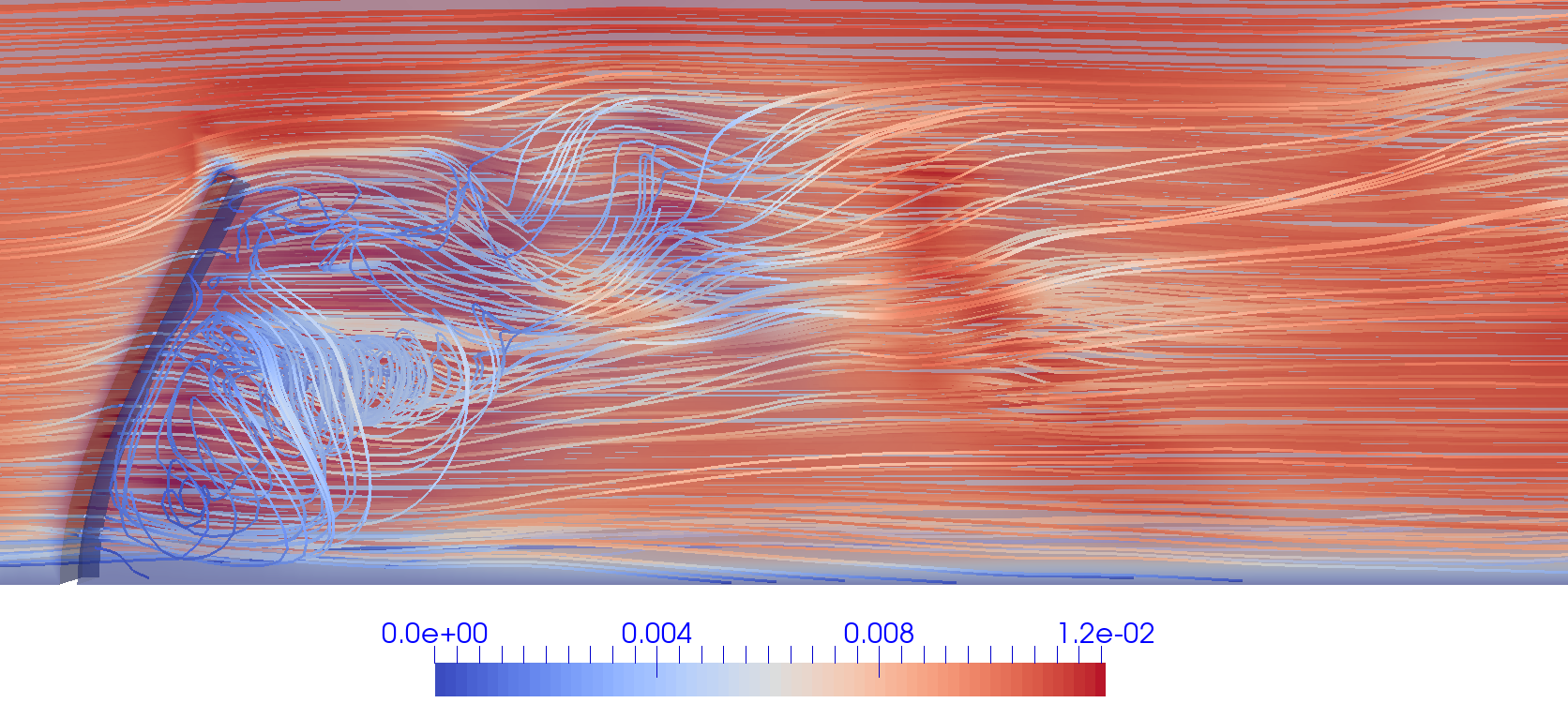}
      \end{overpic}
\begin{overpic}[width=0.47\textwidth]{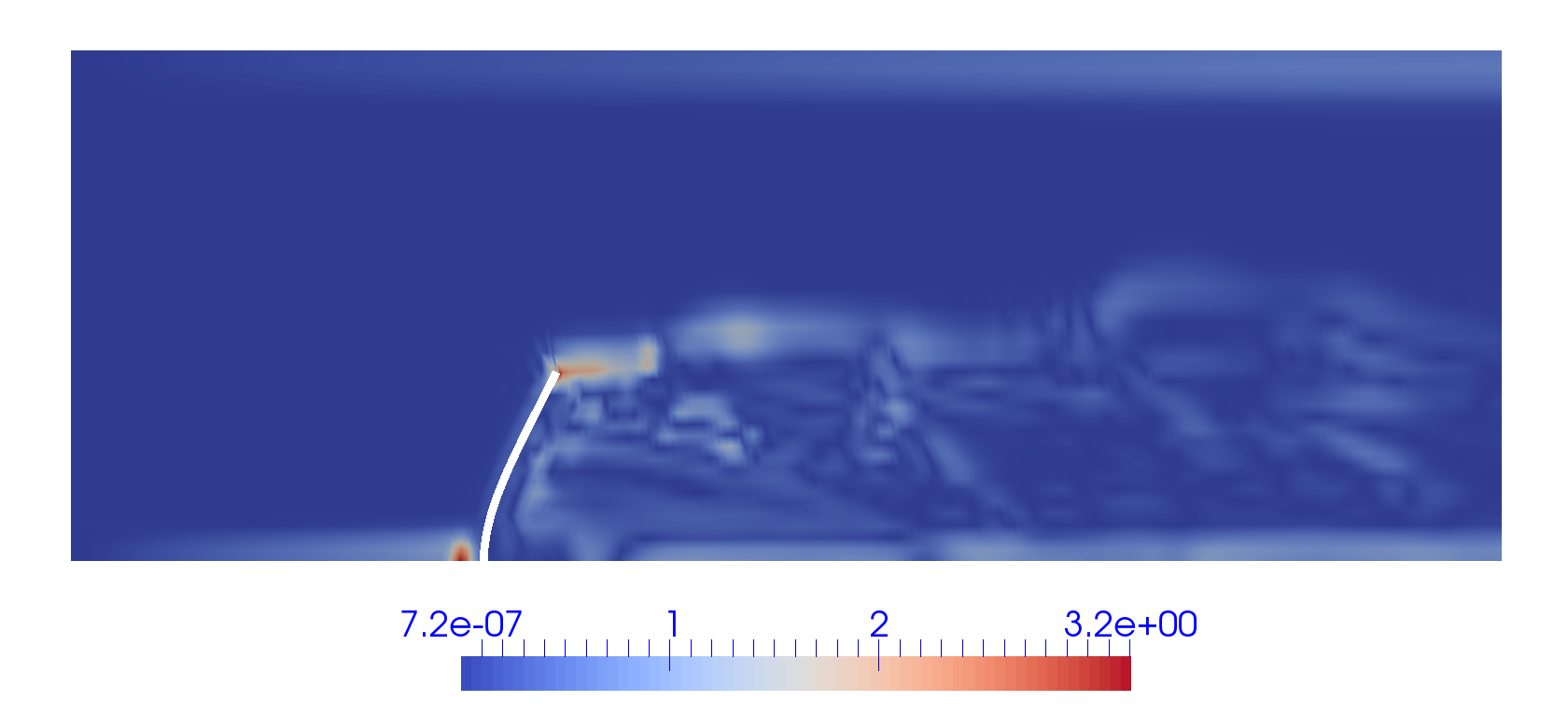}
      \end{overpic}\\
\caption{Mesh $300k$: Velocity $\u_f$ (top left), pressure $p_f$ (top right), streamlines (bottom left) and vorticity (bottom right) fields computed by the NSE algorithm 
close to steady state for Reynolds number 400.}
\label{fig:NSE_2}
\end{figure}

To illustrate the flow field, we show in Figs.~\ref{fig:NSE} and \ref{fig:NSE_2} the velocity, pressure, streamlines and vorticity computed by the NSE
algorithm close to steady state 
for $Re = 100$ and $Re = 400$. The results were obtained with mesh $300k$.
As expected, the flow field becomes more complex as the Reynolds number is
increased.

At the numerical level, for all the simulations we fix the number of PISO loops to 3 and non-orthogonal correctors 
to 1. We use an initial relaxation factor of 0.4 at each time step. 
For the convective term, we adopt a Central Differencing (CD) scheme \cite{Lax1960} 
in order to avoid introducing stabilization associated with upwind schemes. In this way, 
we are able to assess the effectiveness of the differential filter.
At the end of each FSI iteration, we compute the $L^2$-norm of the residual vector on 
the fluid side of the interface: if this norm falls below 1e-6, then 
we stop the iterating between fluid and structure solvers for the given time step. 

We ran all the simulations in parallel using 20 processor cores. 
This means about $15k$ cells per CPU for mesh $300k$ and about $35k$ cells per CPU for mesh $700k$, 
both below the scalability limit for OpenFOAM (i.e., about $50k$ cells per CPU).
The simulations are run on SISSA HPC cluster Ulysses, which 
has recently been upgraded to 200 TFLOPS, 2TB RAM, 7000 cores.
 
We note that the FSI convergence rate decreases when the Reynolds number increases
and during the transition to steady state.

\begin{rem}\label{rem_1}
The release of solids4Foam that we used for the results in this paper 
could not be restarted from a given time step in an efficient way.
This issue, that affects both the serial and the parallel versions, 
is noted also in \cite{Meng2018} for an earlier release. 
The restart introduces a larger error at the FSI interface
with respect to the error of an uninterrupted simulation.
This generates a localized perturbation in the flow field that spoils 
the quantities of interest for the benchmark, in particular the drag coefficient.
We note that by the time this paper was completed, the developers of 
solids4foam had uploaded a fix for this problem affecting the restart. However, since such a fix had not been merged  
with the master branch we chose not to test it.

The restart problem in solids4Foam, together with the maximum wall time of 96 hours allowed for the Ulysses cluster,
limited the level of mesh refinement we could afford. An additional difficulty is given by the long 
time required to reach a quasi steady state condition for the benchmark we consider. 
For all of these reasons, we consider only the lower Reynolds numbers in 
\cite{Tian2014}, and we report only results for the meshes in Table \ref{tab:mesh}. Moreover, for the 
finer computational mesh we could run only the NSE model. 
\end{rem}

At $Re = 100$, the maximum wall time (96 hours of computations) allows NSE to simulate 85 s of flow with mesh $300k$, 
while EFR simulates only 61 s. This means that one time step of NSE (resp., EFR) 
takes in average 40.6 s (resp., 56.6 s). Thus, we estimate that the filter step 
takes in average 16 s per time step with mesh $300k$ at $Re = 100$.
This is in line with the times reported in \cite{BQV}.
Recall, that at each time step the fluid and structure problems are solved multiple times. 
With mesh $700k$, the maximum wall time allows NSE to simulate only 62 s of flow. Considering that the change in the quantities 
of interest for this benchmark slows down after 55 s of flow for $Re = 100$, we 
did not report the results for EFR with mesh $700k$ as they are not close enough to steady state. 
We would like to mention that in \cite{Rege2017} the authors report over 200 computation 
hours with solids4Foam to simulate 2.32 seconds of a FSI problem at $Re \sim 2000$
with a Smagorinski-based LES model. They used  20 Intel Xeon CPUs E7- 8870 @ 2.40 GHz processor cores.

Table \ref{tab:ref} reports the quantities of interest computed in \cite{Tian2014}.
In Table \ref{tab:1}, the NSE solutions obtained with meshes $300k$ and $700k$ and
the EFR solution obtained with mesh $300k$ are compared with the reference values from \cite{Tian2014} 
for $Re = 100$ and $Re = 400$. For the EFR algorithm, we set $\alpha = h_{min}^f$
and the value of $\chi$ has been computed by using the following formula introduced in \cite{Girfoglio2019} for 
flow problems in fixed domain:
\begin{equation}\label{eq:chi}
\chi = \dfrac{h_{min}^f - \eta}{\dfrac{\rho_f \alpha^2}{\mu_f \Delta t} \eta - \eta}. 
\end{equation}
For the current test, we obtain $\chi = 0.028$ at $Re = 100$ and $\chi = 0.0255$ at $Re = 400$. 
From Table \ref{tab:1}, we see that the total error $|E_{c_d}| + |E_{\Delta x/b}| + |E_{\Delta y/b}|$
for the NSE algorithm decreases as the mesh is refined at a given $Re$ and it increases as $Re$ increases,
as one would expect.
In addition, we see that the EFR algorithm performs well both for $Re = 100$ and $Re = 400$: 
it allows to obtain a smaller total error. 
Finally, we note that the NSE algorithm with the finer mesh (i.e., $700k$) does not perform 
as well as the EFR algorithm on the coarser mesh (i.e., $300k$).


\begin{table}[htb!]
\centering
\begin{tabular}{cccc}
\cline{1-4}
$Re$ & $c_d$ & $\Delta x/b$ & $\Delta y/b$ \\
\hline
100 & 1.02 & 2.34  & 0.67 \\
\hline
400 & 0.94 & 2.34  & 0.68  \\
\hline
\end{tabular}
\caption{Drag coefficient and displacement of the structure free edge reported in \cite{Tian2014} for Reynolds numbers 100 and 400.} 
\label{tab:ref}
\end{table}

\begin{table}[htb!]
\centering
\begin{tabular}{ccccccccc}
\hline
\multicolumn{9}{c}{$Re = 100$} \\
\cline{1-9}
Mesh name & Algorithm   & $c_d$ & $\Delta x/b$ & $\Delta y/b$ & $E_{c_d}$ & $E_{\Delta x/b}$ &  $E_{\Delta y/b}$  & $\sum_i |E_i|$ \\
\hline
$300k$ & NSE & 1.24 & 2.235 & 0.625 & 0.216  & -0.045 & -0.067 & 0.328\\
$300k$ & EFR & 1.236 & 2.243  & 0.63 &  0.212 & -0.041 &  -0.06 & 0.313\\
\hline
\hline
$700k$ & NSE & 1.239 & 2.239  & 0.628 & 0.215  & -0.043 & -0.063 & 0.321\\
\hline
\multicolumn{9}{c}{$Re = 400$} \\
\cline{1-9}
Mesh name & Algorithm   & $c_d$ & $\Delta x/b$ & $\Delta y/b$ & $E_{c_d}$ & $E_{\Delta x/b}$ &  $E_{\Delta y/b}$  & $\sum_i |E_i|$ \\
\hline
$300k$ & NSE & 0.958 & 1.809  & 0.402 & 0.019  & -0.227 &  -0.409 & 0.655\\
$300k$ & EFR & 0.963 & 1.842  &  0.418 & 0.024  & -0.213 &  -0.385  & 0.622\\
\hline
\hline
$700k$ & NSE & 1.044 & 1.903  & 0.446 & 0.111  & -0.187 &  -0.344 & 0.642\\
\hline
\end{tabular}
\caption{Quantities of interest and corresponding errors \eqref{eq:Err}
computed with the NSE algorithm (meshes $300k$ and $700k$) and EFR algorithm (mesh $300k$) 
with $\alpha = h_{min}^f$ and $\chi = 0.0255$ (Eq. \eqref{eq:chi}) for $Re = 100$ and $Re = 400$.}
\label{tab:1}
\end{table}


As mentioned above, $\chi$ was set with a formula from \cite{Girfoglio2019}, where
we studied flows at $Re \in [2000, 6500]$. This formula might not be optimized for the Reynolds
numbers we consider here, as it is likely to introduce more artificial dissipation than needed. 
This could explain why in Table \ref{tab:1} the error for the drag coefficient
is much larger at $Re = 100$ than $Re = 400$. Once the restart issue mentioned in Remark \ref{rem_1}
gets fixed, we will work with the higher Reynolds number cases reported in \cite{Tian2014}, 
for which we expect our formula to work better.

\begin{figure}[htb!]
\centering
 \begin{overpic}[width=0.45\textwidth]{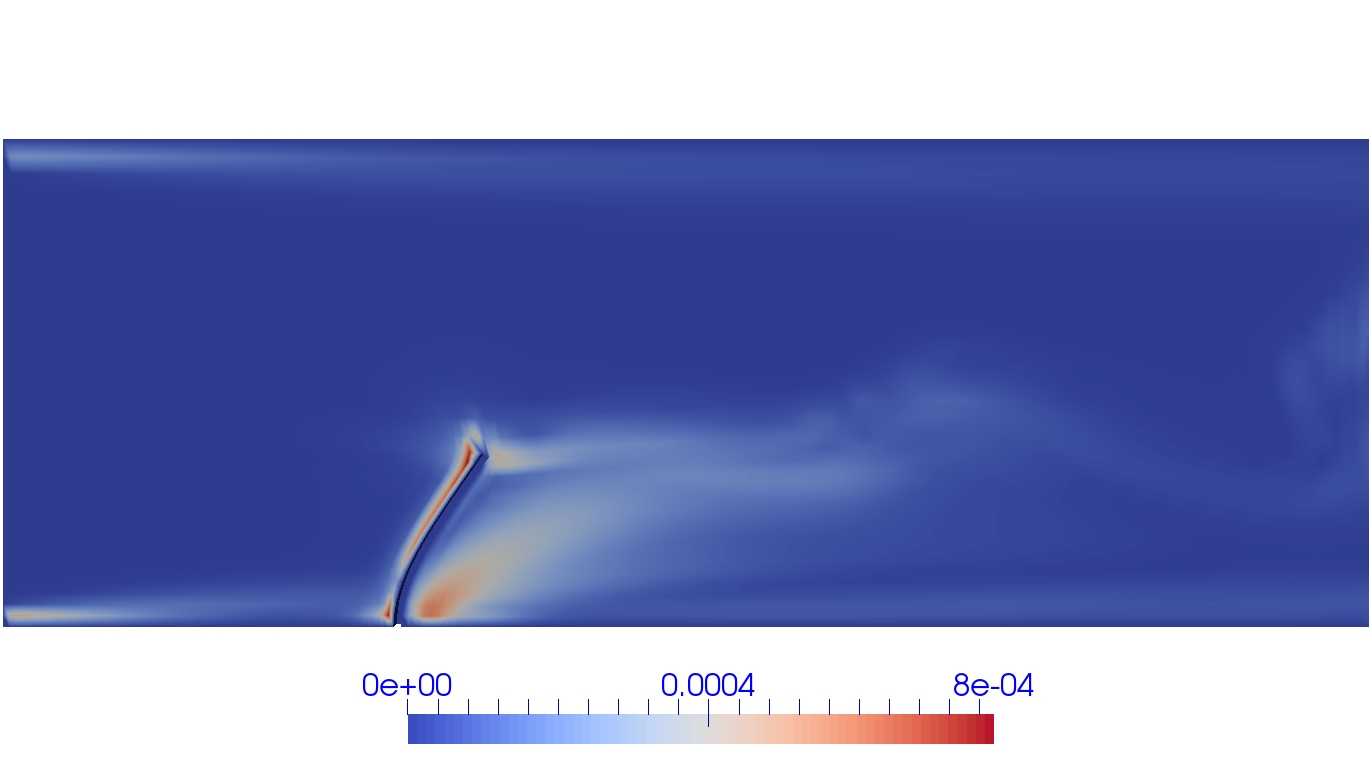}
        \put(40,50){$Re = 100$}
      \end{overpic}
\begin{overpic}[width=0.45\textwidth]{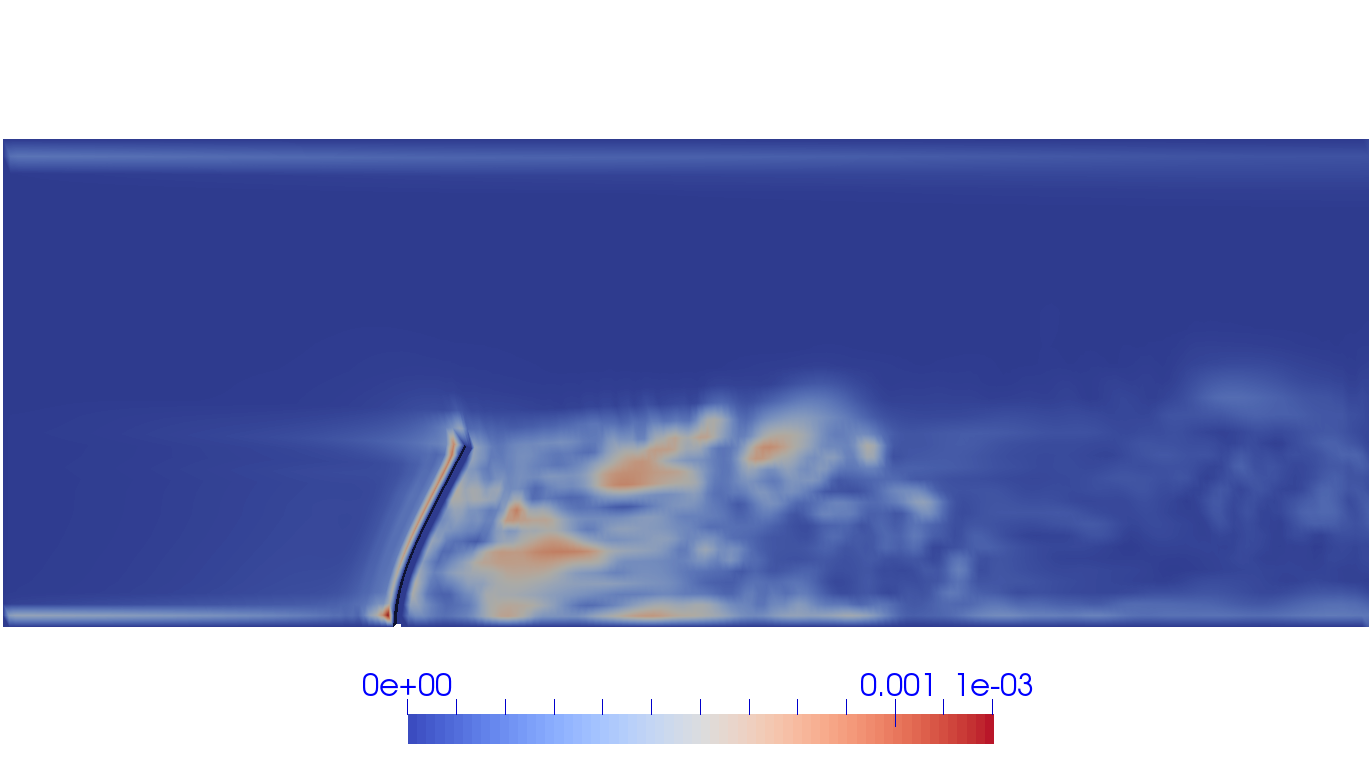}
        \put(40,50){$Re = 400$}
      \end{overpic}\\
\caption{Mesh $300k$: Indicator function computed by the EFR algorithm close to 
steady state for $Re = 100$ (left) and $Re = 400$ (right).}
\label{fig:EFR}
\end{figure}

A key role in the EFR algorithm is played by the indicator function defined in \eqref{eq:a_deconv}.
Fig.~\ref{fig:EFR} shows the indicator function computed with the mesh $300k$ close to steady state. 
We see that the largest values in the region close to and behind the beam, as expected. Additionally, 
local peaks occur inside the channel boundary layer. Thus, function \eqref{eq:a_deconv}
is a suitable indicator function because it correctly selects the regions of the domain where the velocity does need regularization.

\section{Conclusions}\label{sec:conclusions}

This paper has two goals: i) to test open source software solids4Foam, 
which is widely used for FSI simulations; and (ii) 
assess its flexibility in handling more complex flows. To accomplish such goals, 
we considered the EFR implementation of a Leray model to study the interaction of an incompressible
fluid at moderately large Reynolds numbers with a with a hyperelastic
structure modeled as a Saint Venant-Kirchhoff material. We used
a strongly coupled, partitioned FSI solver in a finite volume environment, combined
with an arbitrary Lagrangian-Eulerian approach to deal with the motion of the fluid domain.

With regard to goal i), we found that solids4Foam is accurate when compared against
numerical results in the literature but suffers from a major limitation: the restart function
introduces a larger error at the FSI interface, which spoils the restarted simulation. This limits
the applicability to problems that can be solved in one run, i.e.~simulations
over a short period of time. Thus, even simple benchmark tests like the one studied in this article 
(3D cross flow around an immersed, slender structure \cite{Tian2014}) are feasible only at Reynolds numbers 
of a few hundreds.
It appears that a fix to this bug is forthcoming.
As for goal ii), we found it relatively easy to replace the standard Navier-Stokes
solver with our LES approach within the FSI algorithm. This indicates that solids4Foam is a versatile tool that can easily
be extended to more complex fluid (or structure) models.


\section*{Disclosure statement}
No potential conflict of interest was reported by the author(s).

\section*{Funding}\label{sec:acknowledgements}
This research was funded by the European Research Council
Executive Agency by the Consolidator Grant project AROMA-CFD "Advanced
Reduced Order Methods with Applications in Computational Fluid Dynamics"
- GA 681447, H2020-ERC CoG 2015 AROMA-CFD, PI G. Rozza, INdAM-
GNCS 2019-2020 projects, the US National Science Foundation through grant DMS-1620384 and DMS-195353.

\bibliographystyle{plain}
\bibliography{FSI.bib}





\end{document}